\documentclass{mem}
\usepackage{natbib}\usepackage{txfonts}\usepackage{balance}
\usepackage{graphicx}
\usepackage[a4paper,breaklinks,dvipdfm]{hyperref}
\idline{75}{282}
\begin{document}
\def\teff{$T\rm_{eff }$}
\def\kms{$\mathrm {km s}^{-1}$}
\newcommand{\fermi}{\textit{Fermi}}

\title{
Supernova Remnant Shock - Molecular Cloud Interactions}
   \subtitle{Masers as tracers of hadronic particle acceleration}

\author{
Dale A. \,Frail\inst{1} 
          }

  \offprints{D. A. Frail}

\institute{
National Radio Astronomy Observatory,
Array Operations Center,
P.O. Box O,
1003 Lopezville Road,
Socorro, NM 87801-0387, 
USA. \email{dfrail@nrao.edu}
}

\authorrunning{Frail}

\titlerunning{SNR-MC Interactions}

\abstract{We review the class of galactic supernova remnants which
  show strong interactions with molecular clouds, revealed through
  shock-excited hydroxyl masers. These remnants are preferentially
  found among the known GeV and TeV detections of supernova remnants.
  It has been argued that the masers trace out the sites of hadronic
  particle acceleration. We discuss what is known about the physical
  conditions of these shocked regions and we introduce a potential new
  maser tracer for identifying the sites of cosmic ray acceleration.
  This review includes a reasonably complete bibliography for
  researchers new to the topic of shock-excited masers and supernova
  remnants.

\keywords{ISM: molecules -- masers -- shock waves -- cosmic rays --
  supernova remnants}
}

\maketitle{}

\section{A Supernova Remnant Origin for Cosmic Rays}

Cosmic rays were discovered nearly 100 years ago, yet we still do not
know with any certainty the identity of the cosmic accelerators.  It
has long been argued that supernova remnant (SNR) shocks are the
primary sites for accelerating galactic cosmic rays
\citep[e.g.][]{be87}. Observations of non-thermal radio and X-rays
from SNRs have established that {\it electrons} are accelerated in
these shocks (up to 100 TeV) \citep{kpg+95}, but the bulk of galactic
cosmic rays (99\%) are {\it hadrons} and here solid evidence has been
lacking.

The most promising method of searching for evidence of hadronic cosmic
rays has been to look for GeV and TeV emission from SNRs. The cosmic
ray protons and other hadrons, accelerated in SNR shocks, collide with
the ambient gas producing high-energy particles called neutral pions
which decay quickly into $\gamma$-rays
($\pi^\circ\rightarrow\gamma\gamma$). EGRET provided some early
detections of SNR-$\gamma$-ray associations but with only crude
localizations \citep{ehks96}. Interest in this topic has been
reinvigorated in recent years with a new generation of atmospheric
Cerenkov imaging telescopes (e.g. HESS, MAGIC, VERITAS) and the launch
of NASA's \fermi\ satellite with greatly increased sensitivity and
angular resolution over past experiments.

Simply finding GeV/TeV emission spatially coincident with an SNR is
not sufficient to demonstrate that hadrons are undergoing diffusive
shock acceleration in the SNR. There are other potentially confusing
sources of $\gamma$-ray emission originating primarily from leptonic
(i.e. electron) processes \citep{abks08}.  Most of the identified
galactic TeV sources have turned out to be pulsar wind nebulae (PWN)
where the $\gamma$-rays originate from the pulsar termination wind
shock.  Young SNRs ($\sim10^3$ yrs) can also produce $\gamma$-rays by
Bremsstrahlung emission and inverse Compton scattering of the ambient
radiation field.  Distinguishing between these alternatives is not
straightforward.

SNRs whose cosmic rays are interacting with dense molecular clouds
(MCs) are expected to be bright $\gamma$-ray sources.  In these cases
it is necessary to rule out leptonic processes and to show
quantitative agreement with the predictions of the model in order to
build a strong case for the acceleration of hadronic cosmic rays.  The
$\gamma$-ray flux from neutral pion decay depends on the ratio $n/d^2$
where $d$ is the SNR distance and $n$ is the ambient gas density.
Fortunately radio astronomy provides one simple method to obtain both
of these hard-to-measure parameters. In this review we show that the
OH(1720 MHz) maser transition has become one of the more important
{\it in situ} tracers of molecular shocks in SNRs \citep[see the
excellent review by][]{wy02}.  Through extensive observations and
theory, the OH(1720 MHz) masers have been recognized as a key signpost
for the interaction of SNRs with MCs \citep{jcw+10}.

\section{Observational Properties of OH(1720 MHz) masers}

Anomalous emission from the OH satellite line at 1720 MHz was first
noted toward the SNR W\,44 and W\,28 in the late 1960s \citep{gr68}.
This result was largely forgotten until 25 years later when high
sensitivity, high resolution Very Large Array (VLA) observations
showed 26 compact, narrow line features tracing the interaction zone
between the SNR and the MC \citep{fgs94}. The high brightness
temperature (T$_b>4\times10^{5}$ K) and narrow line widths
($\Delta{\rm V}\simeq 1 $\kms) argued convincingly for a non-thermal,
{i.e.} maser origin for the OH(1720 MHz) emission.

\begin{table*}
\caption{A distance-ordered list of OH(1720 MHz) supernova remnants}
\label{abun}
\begin{center}
\begin{tabular}{lccccccr}
\hline
\\
(l,b) & RA & Dec. & d$_{kpc}$ & \# & $\gamma$-rays? & Name & Reference\\
 & (J2000) & (J2000) & (kpc) & masers &  & & \\ 
\hline
\\
189.1, +3.0  & 06 17 00 &   +22 34 & 1.5 &  6 & Y & IC443 & \citet{cfgg97}\\
6.4,$-$0.1  & 18 00 30 &   -23 26 & 2.0 & 41 & Y & W28 & \citet{cfgg97}\\
34.7,$-$0.4  & 18 56 00 &   +01 22 & 2.5 & 25 & Y & W44 & \citet{cfgg97}\\
5.7,$-$0.0  & 17 59 02 & $-$24 04 & 3.2 & 1 & P &  & \citet{hy09}\\
8.7,$-$0.1  & 18 05 30 & $-$21 26 & 3.9 & 1 & Y & W30 & \citet{hy09}\\
9.7,$-$0.0  & 18 07 22 & $-$20 35 & 4.7 & 1 & N & & \citet{hy09}\\
359.1,$-$0.5 & 17 45 30 & $-$29 57 & 5.0 & 6 & P & & \citet{yur95}\\
5.4,$-$1.2  & 18 02 10 & $-$24 54 & 5.2 & 2 & N & Milne\,56 & \citet{hy09}\\
49.2,$-$0.7  & 19 23 50 &   +14 06 & 6.0 & 2 & Y & W51C & \citet{gfgo97}\\
357.7,+0.3   & 17 38 35 & $-$30 44 & 6.4 & 5 & N & & \citet{ygr+99}\\
357.7,$-$0.1 & 17 40 29 & $-$30 58 & $>$6 &  2 & N & MSH\,17-39 & \citet{fgr+96}\\
348.5,+0.1   & 17 14 06 & $-$38 32 & 8    & 10 & Y & CTB\,37A & \citet{fgr+96}\\
32.8,$-$0.1  & 18 51 25 & $-$00 08 & 5.5/8.5 & 1 & P & Kes\,78 & \citet{kfg+98}\\
0.0,+0.0    & 17 45 44 & $-$29 00 & 8.5  & 28 &   Y & SgrAEast & \citet{psm11}\\
1.0,$-$0.1  & 17 48 30 & $-$28 09 & 8.5  & 1 &   N & & \citet{ygr+99}\\
1.4,$-$0.1  & 17 49 39 & $-$27 46 & 8.5  & 2 &   N & & \citet{ygr+99}\\
31.9,+0.0    & 18 49 25 & $-$00 55 & 9.0  & 2 &   N & 3C\,391 & \citet{fgr+96}\\
337.0,$-$0.1 & 16 35 57 & $-$47 36 & 11   & 3 &   N & CTB\,33 & \citet{fgr+96}\\
21.8,$-$0.6  & 18 32 45 & $-$10 08 & 5.2/11 & 1 & N & Kes\,69 & \citet{gfgo97}\\
346.6,$-$0.2 & 17 10 19 & $-$40 11 & 5.5/11 & 5 & N & & \citet{kfg+98}\\
349.7,+0.2 & 17 17 59 & $-$37 26 & $>$11 & 5 & N & & \citet{fgr+96}\\ 
16.7,+0.1 & 18 20 56 & $-$14 20 & 2.2/14 & 1 & N & & \citet{gfgo97}\\ 
337.8,$-$0.1 & 16 39 01 & $-$46 59 & 12.3 & 1 & P & Kes\,41 & \citet{kfg+98}\\ 
\\ 
\hline
\end{tabular}
\end{center}
\end{table*}

This initial discovery prompted several targeted surveys toward the
known sample of SNRs in our galaxy. The strategy was to carry out a
single dish survey on the NRAO 140-ft, Parkes 64-m, and Green Bank
Telescopes. Candidates were followed up with interferometric imaging
with the Australia Telescope Compact Array and the VLA. The advantage
of this particular approach was that the known galactic SNR population
could be surveyed quickly and efficiently.  The major disadvantage was
the relatively high flux density threshold, suggesting that these
surveys are not complete \citep{hy09}. Future OH maser observations
should be carried out as deep interferometry surveys (S$_{lim}\leq$50
mJy) where there will be less confusion from diffuse, thermal OH
emission.

We list the current sample of 23 SNRs with OH(1720 MHz) maser
detections in Table 1. Successful searches toward the Large Magellanic
Cloud have also been conducted but they are not listed in this table
\citep{bglg04,ry05}. Considering the detections as a whole, about 10\%
of all SNRs have OH(1720 MHz) masers \citep{gfgo97,hy09}. The SNRs
with OH(1720 MHz) masers appear to trace the broad scale distribution
of molecular gas in our galaxy. The detections are not found randomly
throughout the galaxy but rather are found in the inner galaxy, in
particular within the molecular ring and nuclear disk. The luminosity
distribution of the OH(1720 MHz) masers (defined as $S\times{d}^2$,
where $S$ is the peak flux density in mJy and $d$ is the distance in
kpc) is broad, covering 10$^2$ to 10$^6$ mJy kpc$^2$, with a median
$\sim 10^4$ mJy kpc$^2$.

The OH(1720 MHz) masers are located on or near the peaks of the radio
synchrotron emission which is generated by electron acceleration along
the expanding SNR shock.  OH(1720 MHz) masers are preferentially
located along the edges of thin filaments or clumps of molecular gas
with broad lines widths seen ($\Delta{\rm V}\simeq 30$ \kms),
indicative of post-shock gas
\citep{fm98,rr99,lwb+02,shs+04,rrj05,lww+10,zc11}.

The masers in a given SNR all have similar radial velocities with
small scatter. There is close agreement between the velocity of the
masers and the systemic velocity of molecular gas in the vicinity.
This velocity matching has been interpreted to mean that the masers
originate in a shock transverse to the line of sight
\citep{fgr+96,cfgg97}.

OH(1720 MHz) masers also provide a unique way to directly measure the
strength and orientation of the magnetic fields in SNR shocks using
Zeeman splitting of the line \citep{yrg+96,
  cfgg97,kfg+98,bfgt00,hgbc05a,hgbc05b}. These observations have
yielded values for the magnetic field of order 1 mG, giving derived
magnetic pressures comparable to the thermal pressure of the hot gas
interior to the SNR. These same observations give measurements for the
size of the maser spots of order 10$^{15}$ cm or $\sim$100 AU
\citep{cgfd99,hgb+03}.

Two interesting correlations exist for those SNRs with OH(1720 MHz)
masers.  \citet{cfgg97} first noted that the SNRs detected at GeV
energies by EGRET were among the best examples of SNR-MC interactions
(IC\,443, W\,28 and W\,44) and they suggested that OH(1720 MHz) SNRs
would make important candidates for future GeV and TeV surveys. The
new generation of instruments supports this initial supposition, with
several new claims of associations between SNRs and $\gamma$-ray sources
\citep{hyw09,cs10,ko10}. We list these associations in Table 1 (Y=yes,
N=no, P=possible).

\citet{fgr+96} and \citet{gfgo97} first noted that the OH(1720 MHz)
SNRs belonged predominantely to a particular class of so-called
``mixed morphology'' SNRs (25\% of X-ray SNRs) which have
center-filled thermal X-ray emission.  This hypothesis was put on a
firmer statistical footing by \citet{ywrs03}, and while there is no
agreement on how this interior X-ray gas is produced
\citep{csm+99,wl91}, the strong correlation argues that SNR-MC
interactions play an important role.

\section{The origin and excitation of OH(1720 MHz) masers}

The hydroxyl (OH) molecule is abundant in the interstellar medium.  In
its ground state it has four ground state transitions: two main lines
at 1665.4 MHz and 1667.4 MHz, and two satellite lines at 1612.2 MHz
and 1720.5 MHz \citep{eli92}. The 1665 and 1667 MHz lines are
typically associated with star-forming regions while 1612 MHz are
associated with evolved stars, but all of these lines are inverted
through pumping of far-infrared photons. OH(1720 MHz) masers, in
contrast, are pumped through collisions; far-infrared radiation
effectively acts to destroys the inversion of the 1720 MHz line.

Theoretical modeling suggests that the OH is formed downstream of a
slow, compression-type shock (20-30 \kms) that has propagated into an
adjacent MC. A strong OH(1720 MHz) maser inversion is collisional
excited at temperatures of 30-120 K and densities of order
$n=$10$^4$-10$^{5}$ cm$^{-3}$ \citep{lge99,war99}. Maser amplification
needs large column densities of OH molecules (10$^{16}-10^{17}$
cm$^{-2}$) with {\it small} velocity gradients.  Thus masers will
occur preferentially where the observer's line-of-sight velocity
gradient is small. OH(1720 MHz) masers therefore favor edge-on
geometries (i.e. transverse shocks). This immediately explains why the
observed velocity of the masers, SNR and MC agree.  Thus the detection
of an OH(1720 MHz) maser is not just unambiguous proof that an SNR-MC
interaction is taking place, it also provides both a kinematic
distance $d$ and the gas density $n$. The ratio $n/d^2$ is a key
parameter for testing the hadronic acceleration models \citep{abks08}.

The ionization of the OH remains an interesting area for future
research. Most of the gaseous oxygen in the dense, post-shock gas is
taken up in H$_2$O formation and therefore it must be dissociated to
get the required columns of OH. \citet{war99} produced a
self-consistent physical model in which the thermal (${\rm T}<$ 1 keV)
X-rays in the interior provide the necessary ionization, conveniently
explaining the association between OH(1720 MHz) SNRs and mixed
morphology SNRs \citep{ywrs03}. More recently \citet{hyw09} have
suggested that the H$_2$O could be dissociating instead from the local
cosmic rays accelerated in the shock. If true, OH(1720 MHz) masers are
not merely a signpost of hadronic particle acceleration; rather, the
locally enhanced cosmic ray density is necessary to produce the
OH(1720 MHz) masers. This hypothesis can be tested directly with
observations of the cosmic ray ionization rate $\zeta_{\rm CR}$ using
the abundance ratio of [DCO$^+$]/[HCO$^+$] \citep{chm+11} in dense
clouds or measuring H$^+_3$ columns \citep{ibg+10} in diffuse clouds.

\section{A new maser tracer for molecular shocks}

OH(1720 MHz) masers are not infallible tracers of SNR-MC interactions.
Only 10\% of the known SNRs in our galaxy have OH masers and GeV/TeV
emission is not seen in about half of the OH(1720 MHz) SNRs. Due the
narrow physical conditions needed to excite the OH(1720 MHz), the {\it
  absence} of masers tells us nothing about whether hadronic cosmic
rays could be responsible for the observed $\gamma$-rays from an SNR.

OH(1720 MHz) is not the only shock-excited maser. Theoretical
predictions suggest that excited-state OH may be present under similar
physical conditions. However, searches for these lines near 6 GHz have
not yielded any detections \citep{fsp07,pfs+08,mwv08} nor have any 22
GHz water masers been found near SNRs \citep{cgfs99,wg07}.
Additionally, the widespread methanol molecule (CH$_3$OH) can be
collisionally pumped, resulting in not just one line but several dozen
bright maser transitions at radio wavelengths. Theoretical modeling
shows that the brightest shock-excited masers are expected for the
transitions at 36.169 GHz and at 44.070 GHz, with slightly weaker
masers for the 84.521 GHz and 95.169 GHz transitions
\citep{mko85,cjgb92}. These CH$_3$OH masers are excited over a much
larger range of densities and temperatures than OH(1720 MHz).

The first detection of a CH$_3$OH maser toward an SNR was reported by
\citet{zs08} toward Kes 79 at 95 GHz. Observations at 95 GHz with the
Arizona Radio Observatory 12-m and VLA observations at 44 GHz failed
to confirm this detection (Claussen and Frail, unpublished). Methanol
masers have been detected at 36 and 44 GHz toward the galactic center
\citep{hmb90,shhb89}, while recent VLA observations have localized
them to where the SNR Sgr A East is seen interacting with a molecular
ridge \citep{spf10,psf11}. There is evidence that the CH$_3$OH masers
are offset from the known OH(1720 MHz) masers in this region and that
the CH$_3$OH may be excited in hotter and denser gas than the OH (i.e.
closer to the molecular shock front).  Motivated by this intriguing
result, we have recently undertaken a VLA CH$_3$OH survey (36 and 44
GHz) of 21 OH(1720 MHz) SNRs (Pihlstr\"om et al. in preparation).
Owing to the small field-of view (1-arcmin at 44 GHz vs 27-arcmin at
1720 MHz) the survey is strongly biased. We have focused our VLA
observations mainly toward OH(1720 MHz) masers sites and toward some
GeV/TeV peaks.

In our partial reduction of the survey data we have detected a 44 GHz
CH$_3$OH maser in the SNR W\,28. The maser is found in region OH-D,
well away from the bulk of the known OH(1720 MHz) masers. More
intriguingly, the CH$_3$OH maser lies near the peak in the TeV
emission of W\,28 \citep{aab+08} where no OH(1720 MHz) had been seen
previously. Pumping models predict that this transition is excited at
temperatures of T=80-200 K and densities $n=10^5-10^6$ cm$^{-3}$,
conditions both hotter and more dense than that needed to excite the
OH(1720 MHz) line. This preliminary result potentially makes methanol
a new and important signpost for SNR-MC interactions, giving us a more
robust and ubiquitous tracer to help us pinpoint SNR-MC interactions
and measure their physical properties over a wider range.

\begin{acknowledgements}
  I thank my collaborators Ylva Pihlstr\"om, Mark Claussen and Lor\'ant
  Sjouwerman for allowing me to discuss our preliminary methanol
  results. The Very Large Array is operated by the National Radio
  Astronomy Observatory, a facility of the National Science Foundation
  operated under cooperative agreement by Associated Universities,
  Inc.
\end{acknowledgements}

\bibliographystyle{aa}

\end{document}